\documentclass[onecolumn]{article}

\usepackage{graphicx}
\usepackage[top=2cm, bottom=2cm, left=2cm, right=2cm]{geometry}
\usepackage[textsize=footnotesize]{todonotes} 
\usepackage{upgreek}
\usepackage{authblk}
\usepackage[font={small},labelfont=bf]{caption}

\usepackage[square,super,comma,numbers,sort&compress]{natbib}
\bibpunct{}{}{,}{s}{}{\textsuperscript{,}}

\title{\textbf\noindent{A novel method for the injection and manipulation of magnetic charge states in nanostructures}}

\renewcommand{\deg}{$^{\circ}$ }

\begin{document}

\author{J. C. Gartside\thanks{j.carter-gartside13@imperial.ac.uk}}
\author{D. M. Burn\thanks{david.burn@diamond.ac.uk}}
\author{L. F. Cohen}
\author{W. R. Branford}
\affil{Blackett Laboratory, Imperial College London}

\renewcommand\Authands{ and }

\maketitle

\begin{abstract}

\noindent Realising the promise of next-generation magnetic nanotechnologies is contingent on the development of novel methods for controlling magnetic states at the nanoscale. There is currently demand for simple and flexible techniques to access exotic magnetisation states without convoluted fabrication and application processes. 360\deg domain walls (metastable twists in magnetisation separating two domains with parallel magnetisation) are one such state, which is currently of great interest in data storage and magnonics. Here, we demonstrate a straightforward and powerful process whereby a moving magnetic charge, provided experimentally by a magnetic force microscope tip, can write and manipulate magnetic charge states in ferromagnetic nanowires. The method is applicable to a wide range of nanowire architectures with considerable benefits over existing techniques. We confirm the method's efficacy via the injection and spatial manipulation of 360\deg domain walls in Py and Co nanowires. Experimental results are supported by micromagnetic simulations of the tip-nanowire interaction.
\\
\end{abstract}

\noindent Magnetic nanostructures can be characterised by maps of their magnetic charge distribution. Magnetic charges arise from a combination of geometric patterning \cite{chou-charge} and internal micromagnetic structure such as that found in magnetic domain walls (DWs) \cite{allwood-charge-walls,PhysRevB.81.020410}. Gaining a thorough understanding of magnetic charge distributions in nanostructures and techniques for their control is essential for next-generation magnetic technologies. In particular, it is important to develop powerful and flexible methods for the manipulation and control of magnetic charge. Existing methods for the injection of magnetic charge into nanostructures rely on complex solenoid-based write heads \cite{terabit-write-head}, subjecting whole devices containing injection pads to global field sequences \cite{lacour2015localization, APL_100_062407, geng2012generation, hehn2008360, IEEE.Trans.Magn_40_2655, J.Korean.Phys.Soc_63_441} or using complex nanowire geometries to locally apply pulsed Oersted fields \cite{muratov-storage,APL_103_222404,JAP_115_17D135,thomas2012topological}. These methods are non-trivial in both fabrication and field-protocol stages and other than solenoidal write-heads inflexible in that they demand spatially fixed charge injection points, predetermined at the device design stage.

Here we report a novel method for the injection and control of magnetic charge in nanostructured magnetic material using the stray field of an external moving magnetic charge. The moving charge is provided here experimentally by a magnetic force microscopy (MFM) tip and allows for the flexibility in spatial injection point whilst providing highly precise, local injection of magnetic charges over a wide range of nanowire geometries, in addition to the spatial manipulation of existing charge structures. To confirm the viability of the method we demonstrate the injection and subsequent manipulation of a bound magnetic charge pair composed of two transverse DWs, known as a 360\deg DW \cite{belavin1975metastable}. 360\deg DWs are currently of great interest in data storage and magnonics, as well as intriguing topological defects in their own right. They have been proposed as candidates for high-density data storage \cite{benitez2015magnetic} including magnetoresistive random access memory (MRAM) systems \cite{muratov-storage,PhysRevB.87.214403,multi-level-storage} and as phase-shifting \cite{hertel2004domain} and frequency doubling \cite{mascaro2010ac} magnonic circuit elements as well as spin-wave generators \cite{mascaro2010ac}. Experimental progress of such applications has been so far hampered by a lack of simple and versatile means to access 360\deg DW states, an issue which this method addresses. Additionally,  we demonstrate an elegant extension of the injection technique to achieve the controlled motion of 360\deg DWs through nanowires. Spatial manipulation of 360\deg DWs is not possible using uniform global fields as opposite forces are generated on each composite magnetic charge \cite{ross-field-collapse,muratov-field-collapse,kunz2009field} and current-driven motion requires high current densities \cite{mascaro2010ac,dong-jingguo-current-motion} or multiple-wire geometries \cite{zhang2014faster} as well as the incorporation of macroscopic electrodes within the device architecture. A tip-based technique for writing magnetic charge structures into continuous films has recently been described \cite{albisetti2016nanopatterning}, relying on a heated tip in conjunction with a complex multi-layered film and requiring a considerable applied global magnetic field to function. Here we provide a simple and versatile alternative to existing techniques. While this work was under review, a methodology for reconfiguring a nanopatterned magnetic charge array using an MFM tip along with global magnetic fields was described \cite{wang2016rewritable}. The technique is essentially global field mediated with a localised field enhancement provided by an MFM tip raised a distance above the sample. The technique described in this work requires no global field and as such offers considerable benefits.

\section*{Results}

\begin{figure}[tbp]
	\centering
	\includegraphics[width=14cm]{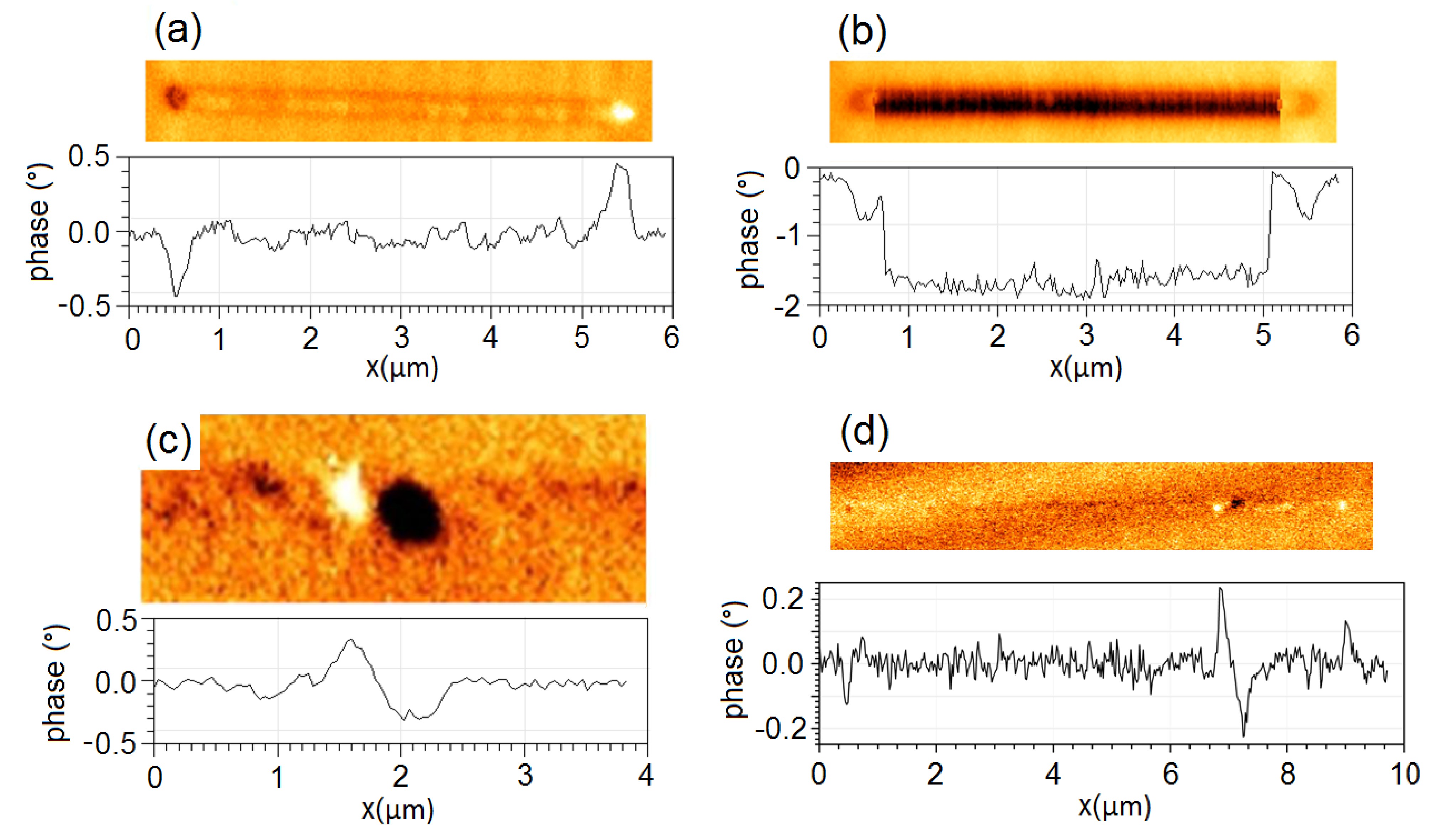}
	\caption{MFM images of Co nanowires in imaging (LM) and injection (HM) modes. Dark contrast marks regions where a negative phase shift was required to maintain constant phase in the cantilever oscillation. This corresponds to an attractive tip-sample interaction, with light contrast indicating a repulsive interaction.
\textbf{a)}	LM-MFM image of a 5 $\upmu$m nanowire magnetised in-plane. No DWs are present.
\textbf{b)} HM-MFM image of same 5 $\upmu$m nanowire. Dark contrast throughout the wire indicates DW injection.
\textbf{c)} LM-MFM image taken after HM injection process. Adjacent regions of opposing magnetic charge are observed, corresponding to an injected 360\deg DW.
\textbf{d)} LM-MFM image after HM injection process showing a whole nanowire containing a 360\deg DW. Wire end charges are seen to have half the magnitude of the injected DW. \textbf{a)}-\textbf{c)} show 150 nm wide nanowires, \textbf{d)} shows a 100 nm wide wire. The HM injection scan leading to the state shown in \textbf{d)} was performed in a contact mode with no oscillation of the tip-height while the injection scan leading to the state shown in \textbf{c)} was performed in a tapping mode with resonant oscillation of the tip-height, no difference is observed between the resultant 360\deg DW states. Further discussion of tapping and contact modes is provided in the methods section.}
\label{mfm_inj-series}
\end{figure}

\subsection*{Experimental demonstration of the injection process}

Initially the nanowires were saturated using a global magnetic field applied along their axis. 
Figure \ref{mfm_inj-series} a) shows a low-moment (LM) tip MFM scan of a typical nanowire (here Co) after the application of this global field. The dark and light contrast at the ends of the wire show the location of the magnetic charges and in this case confirm the expected single-domain state prior to injection. 

Figure \ref{mfm_inj-series} b) shows an MFM scan of the same nanowire performed using a high-moment (HM) tip. In this case the dark contrast throughout the wire shows the tip is attracted to the wire, suggesting that a magnetic charge is present under the tip at all points of the scan along the nanowire. A line profile taken along the wire is shown below the MFM image. The continuous contrast throughout the centre of the wire is seen to have twice the magnitude of the end charges. As DWs possess twice the magnetic charge of a geometrical wire-end charge, this suggests a DW has been introduced to the wire (the central portion of the wire was confirmed to be initially charge free by a prior LM-MFM scan). After the HM scan, the tip is retracted at the midpoint of the wire.
In figure  \ref{mfm_inj-series} c) a subsequent LM-MFM scan is performed, showing a pair of opposing polarity magnetic charges present in the wire. Again, these are each twice the magnitude of the end charges, indicating head-to-head and tail-to-tail 180\deg DWs in close proximity. The two 180\deg DWs must have the necessary opposing chiralities to form a stable 360\deg DW state as oppositely charged DWs of matching chirality will simply attract and annihilate\cite{kunz2009field}. Figure \ref{mfm_inj-series} d) shows an entire nanowire with an injected 360\deg DW. The constituent 180\deg DWs can clearly be seen to possess twice the magnitude of the end charges and no other magnetic charges are present in the wire.   

These three stages demonstrate experimental confirmation of the injection process: LM-MFM imaging of the wire prior to injection, an HM-MFM scan to inject magnetic charges and a post-injection LM-MFM imaging of the wires showing the injected charge structure. It was found that allowing the HM tip to perform subsequent scan lines in a raster fashion did not lead to the injection of additional 360$^{\circ}$ DWs with each line. This has the benefit that 360$^{\circ}$ DWs may be precisely placed by allowing the tip to raster scan until the desired injection location is reached at which point the HM-tip is retracted, leaving a single 360$^{\circ}$ DW. The scan shown in figure \ref{mfm_inj-series} c) was performed at the retraction point of a previous HM scan. Micromagnetic simulations of scans on wires containing an existing 360$^{\circ}$ DW show the tip field causing the DW to collapse before the tip reaches the wire (discussed further in the simulation section below). The tip then injects a new 360$^{\circ}$ DW as it crosses the wire, leaving a single 360$^{\circ}$ DW in the wire as observed experimentally. A video of the time evolution of this process is included in the supplementary materials. 

\subsection*{Micromagnetic simulation of the injection process}

\begin{figure}[tbp]
	\centering
	\includegraphics[width=12cm]{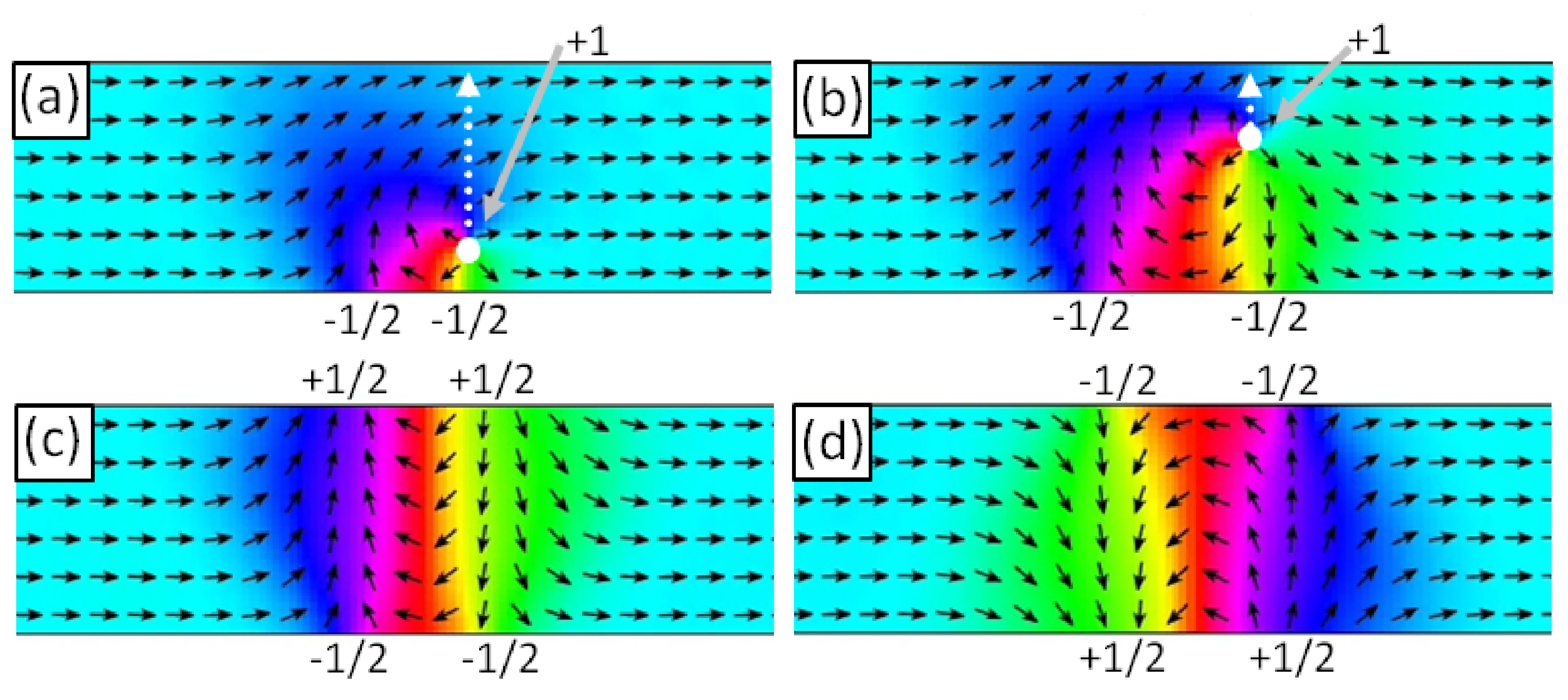}
	\caption{The micromagnetic structure of a nanowire distorted by the movement of a magnetic charge. Topological defects present are labelled with their winding numbers. The moving magnetic charge and its direction of motion are represented by the white arrow and dotted line respectively. The wire simulated here is 150 nm wide by 10 nm thick, with the material properties of Py. \textbf{(a)} and \textbf{(b)} show how the arrangement of topological defects within a nanowire can be manipulated by the passing charge. \textbf{(c)} and \textbf{(d)} show the final micromagnetic structures of 360\deg DWs with opposing chirality, resulting from the charge moving across the wire in up and down directions respectively. An accompanying video is included in the supplementary information.}
	\label{360DW_topology}
\end{figure}

Micromagnetic simulations provide further insight into the physical behaviour taking place whilst the field from a monopole-like tip interacts with the nanowire structure. Figure \ref{360DW_topology} shows a series of micromagnetic configurations of a nanowire as (a) and (b) a magnetic charge passes over the nanowire and (c) and (d) 360\deg DW structures stabilise in the nanowire following the interaction with the moving magnetic charge. Videos of the time evolution of this process are included in the supplementary materials.

We find that the MFM tip locally distorts spins in the nanowire out of the homogenous ferromagnetic state and into a conformation following that of the monopole-like tip field as described by Magiera et al\cite{magiera2014magnetic,magiera2012magnetic}. This introduces a magnetic vortex-like structure into the magnetisation texture which stabilises into a 360$^{\circ}$ DW once the tip has crossed the wire. 

It is perhaps clearest to describe the injection process in terms of dynamic topological defects \cite{mermin-topology}, both those with integer winding numbers which are free to move within the wire and those with fractional winding numbers, bound to the wire edges. Topological defects in thin-film ferromagnets are points at which spins diverge from a uniform collinear texture in a manner which cannot be smoothly unwound. Each defect has an associated winding number describing the manner in which spins locally diverge. The net winding number of a system is rigorously conserved and in a nanowire must sum to zero\cite{PRL_95_197204}. In figure \ref{360DW_topology} a) the monopole-like field of the moving magnetic charge (illustrated by the solid white circle) distorts the uniformly magnetised micromagnetic structure of the nanowire. A spin rearrangement to lower the Zeeman energy in the vicinity of the magnetic charge results in the formation of a topological defect with a $+1$ winding number \cite{magiera2014magnetic,magiera2012magnetic} directly under the moving charge. The introduction of this defect at the wire's lower edge is accompanied by the formation of two additional topological edge defects with $-1/2$ winding numbers in order to conserve the net winding number of the wire. These fractional defects remain bound to the lower edge of the wire, this can be understood by examining the energetics as their leaving the edge would create a discontinuity in the magnetisation along a line between the wire edge and the defect with a large associated energy penalty\cite{PRL_95_197204}. As the magnetic charge moves across the nanowire, figure \ref{360DW_topology} b) shows the $+1$ defect following the movement of the charge. On reaching the upper edge of the wire the $+1$ defect can no longer follow the motion of the magnetic charge and separates into two $+1/2$ topological defects, bound to the wire edge as seen in figure \ref{360DW_topology} c). Lines of spins oriented perpendicular to wire's length are seen connecting each $-1/2$ defect to the corresponding $+1/2$ defect on the opposite edge of the wire. Each line and connected pair of defects represents a 180\deg DW, hence the resultant micromagnetic structure created by the injection process is a bound state of two 180\deg DWs forming a stable 360\deg DW, here with an anti-clockwise chirality. The movement of the tip across the wire in the downwards direction results in the injection of a 360\deg DW with reversed chirality. The final state of this process is shown in figure \ref{360DW_topology} d). The sign on the topological defects is inverted with respect to c) and the magnetisation now rotates in a clockwise direction. As such our method provides chirality-selective injection, key to controlling DW dynamics in nanowire-network systems \cite{pushp2013domain,zeissler2013non}.

Simulations were performed here with the moving magnetic charge at a constant height above the nanowire. Experimental injection scans were performed with the tip in both a non-tapping contact mode and in tapping mode. The contact mode simplifies the dynamics of the injection process and better corresponds to simulation. Figure \ref{mfm_inj-series} d) shows a 360 \deg DW state injected using a contact-mode injection scan. However, to improve imaging quality and reduce tip damage most injection scans were performed in a tapping mode where the tip height is constantly varied by oscillating the MFM tip cantilever at its resonant frequency. Figure \ref{mfm_inj-series} c) shows a 360 \deg DW injected using a tapping-mode injection scan. No difference in the resultant 360 \deg DW state was observed between tapping and contact modes.

\begin{figure}[tbp]
	\centering
	\includegraphics[width=12cm]{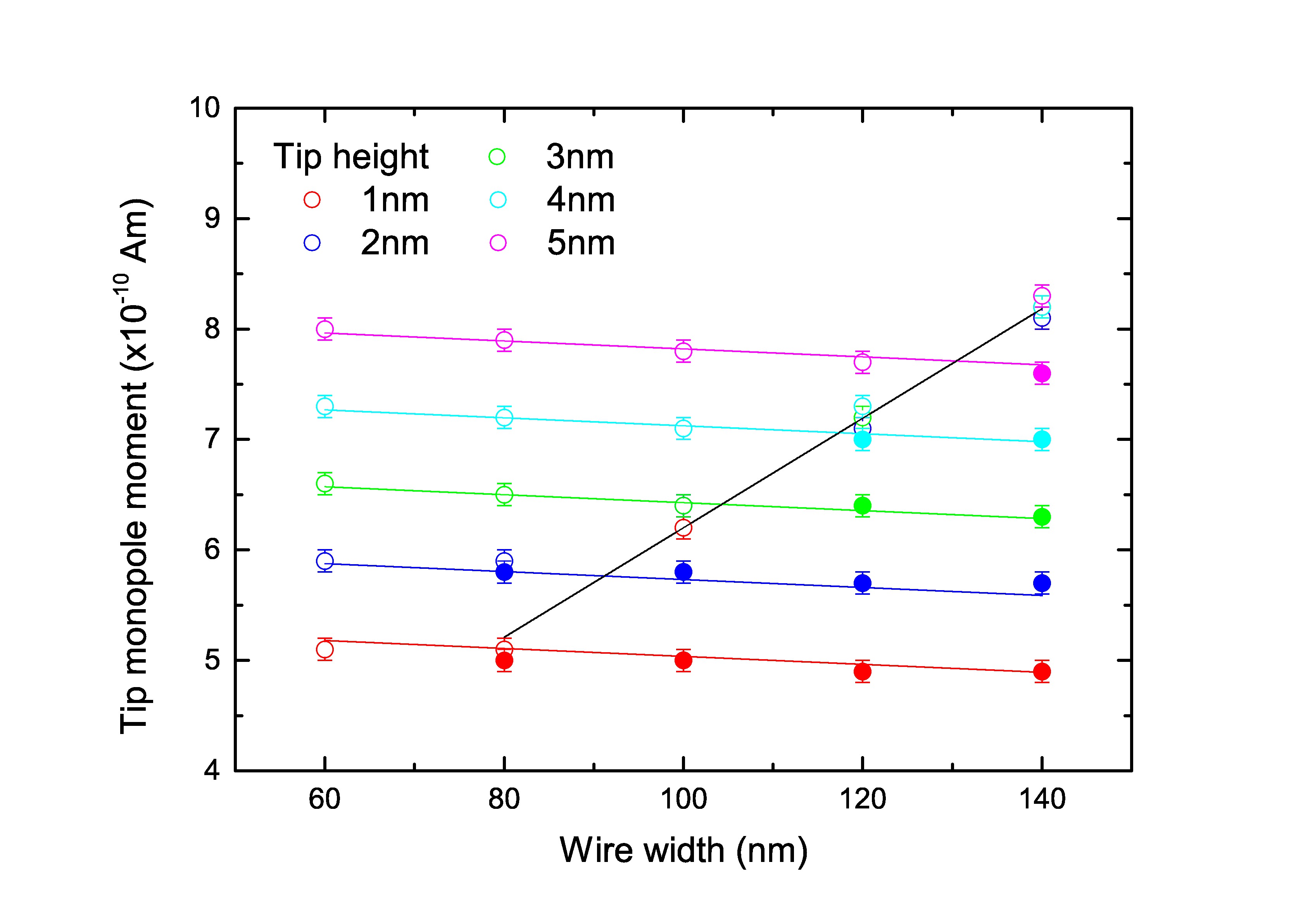}
	\caption{Magnetic charge magnitude required to inject a 360\deg DW into a 10 nm thick Py nanowire. Open points represent successful injection of a stable 360\deg DW structure whilst closed points indicate unstable 360\deg DW structures which collapse once the moving charge has passed the wire. The critical tip charge strength required for stable injection is denoted by the black line.}
	\label{oommf_nucleation}
\end{figure}

The dependence of the nucleation process upon the properties of the MFM tip was investigated through further simulation. Figure \ref{oommf_nucleation} plots the final micromagnetic state of the nanowire following the pass of the magnetic charge for a range of nanowire widths, magnetic charge magnitudes and charge-wire height separations. 

For small magnetic charges, the injection of a 360\deg DW does not occur but above a critical charge magnitude a 360\deg DW is injected. This critical strength shows a gradual increase with increasing nanowire width and a more significant increase with greater separation between the nanowire and magnetic charge. The lines in figure \ref{oommf_nucleation} show a fit to the data using the simple phenomenological form $P_{injection} = \alpha h + \beta w + \gamma$ where $P$ is the critical charge strength, $h$ the height of the charge above the wire and $w$ the wire width. Here $\alpha = 69.6 \pm 0.8$~A, $\beta = -360 \pm 40$~$\upmu$A and $\gamma = 4.70 \pm 0.05$~Am. 

For the 360\deg DW to remain stable in the nanowire following the injection process the magnitude of the magnetic charge must be greater than a secondary critical value. This is represented by the black line on figure \ref{oommf_nucleation}, which increases with nanowire width but as expected shows no magnetic charge / nanowire separation dependence. This dependence is of the form $P_{stable} = \beta w + \gamma$ where $\beta = 5.0 \pm 0.2$~A and $\gamma = 1.2 \pm 0.2$~Am. 

\begin{figure}[tbp]
	\centering
	\includegraphics[width=13cm]{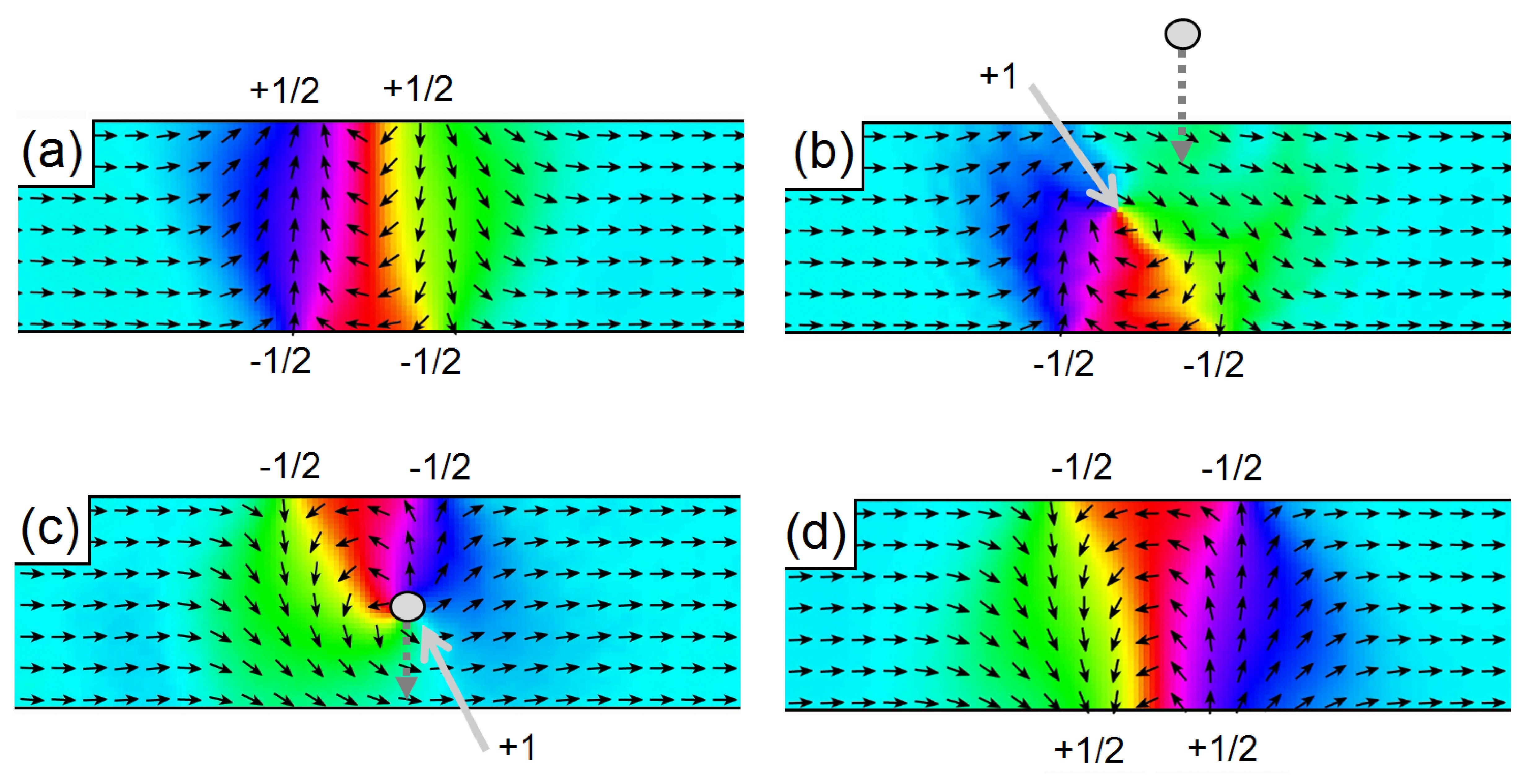}
	\caption{An existing 360\deg DW is destroyed by an approaching magnetic charge before a new 360\deg DW is injected. Topological defects are labelled with their winding numbers, the moving charge and its direction of motion are represented by the grey circle and dotted arrow respectively. \textbf{a)} shows an initial stable 360\deg DW which is caused to collapse by an approaching magnetic charge in \textbf{b)}. The charge then crosses a DW-free wire \textbf{c)}, effecting the same injection process depicted in figure \ref{360DW_topology} and leaving a new stable 360\deg DW \textbf{d)} after passing the wire. An accompanying video is included in the supplementary information.}
	\label{oommf_raster}
\end{figure}

To investigate why each raster line of the HM-MFM injection scan does not inject an additional 360\deg DW an injection process was simulated with an existing 360\deg DW located at the point where the tip-charge crosses the wire. Figure \ref{oommf_raster} a) shows a nanowire containing a stable 360\deg DW. When a moving magnetic charge approaches the wire, its local field forces the spins forming the fractional edge defects on the near edge of the wire to align with the monopole-like field, the reduction in Zeeman energy overcoming the potential binding the fractional defects to the wire edge. The fractional defects then combine into an integer vortex defect, shown in figure \ref{oommf_raster} b). The vortex defect is forced across the wire as the spins around it relax to a collinear ferromagnetic state, lowering their exchange energy. Upon reaching the far side of the wire the vortex meets and annihilates the opposite polarity fractional defects bound to the far edge, leaving no DW in the wire. This process occurs before the moving magnetic charge reaches the wire, hence the charge encounters a wire in a collinear spin state and a new injection process occurs as normal c), leaving a single stable 360\deg DW in the wire d) after crossing. The tip-mediated collapse of existing DWs presents itself as a DW deletion method if the tip is halted after collapsing the existing DW, but before reaching the wire.

\begin{figure}[tbp]
	\centering
	\includegraphics[width=14cm]{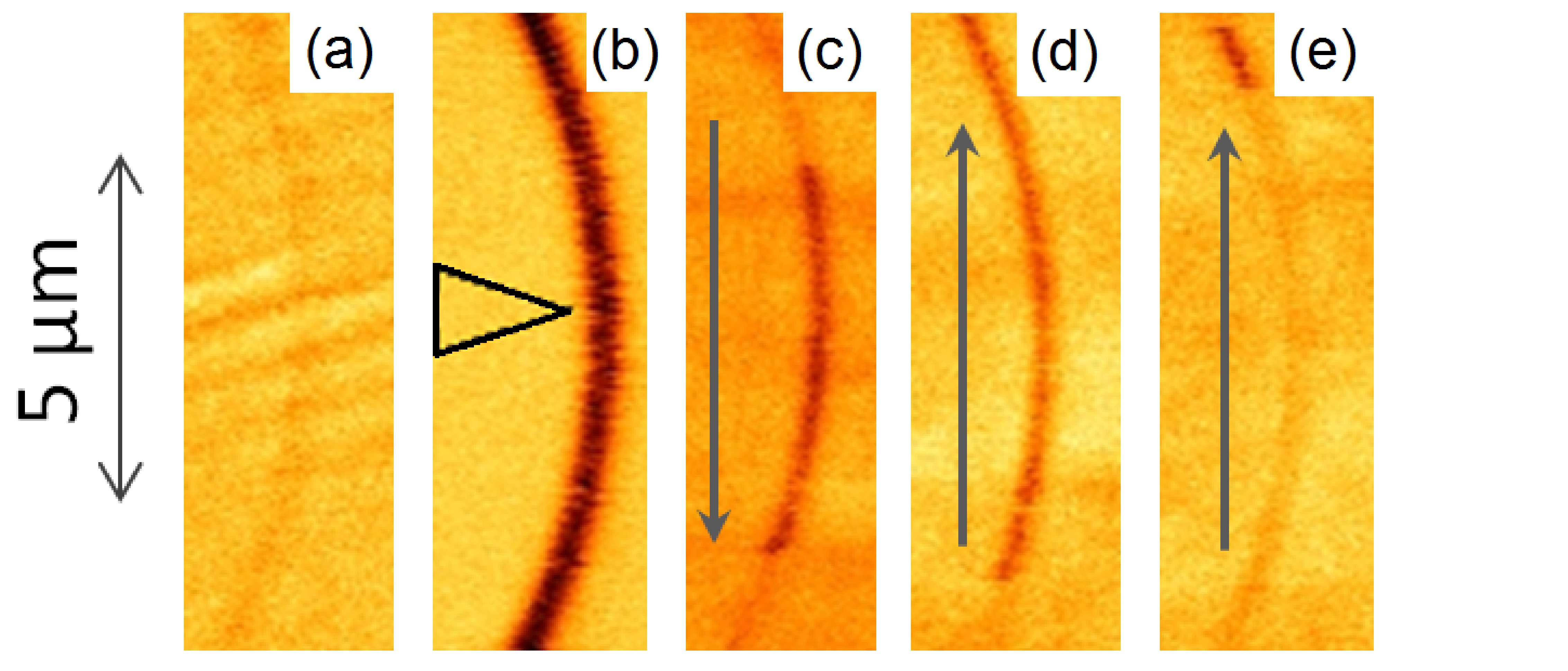}
	\caption{Series of MFM images illustrating 360$^{\circ}$ DW injection and spatial manipulation in a Py nanowire. The scale bar to the left corresponds to all images. \textbf{b)} uses an HM-tip, other images LM. The black triangle in \textbf{b)} indicates the point of tip retraction. Prior to retraction, the wire was raster scanned from the top to the bottom of the image frame to collect the image data before scanning half-way back up the frame and retracting at the point marked by the black triangle. The grey arrows in \textbf{c)}-\textbf{e)} indicate the slow-axis of the raster scan direction. Scans were performed in chronological order from \textbf{a)}-\textbf{e)}. Wire curvature is not a necessary condition for 360\deg DW injection and manipulation, the same dynamics are observed in straight wires.}
	\label{mfm-DW-drag}
\end{figure}

\subsection*{Experimental demonstration of spatial manipulation}

While the weaker magnetic charge associated with a low moment tip is not sufficient to nucleate 360\deg DWs, it can still influence pre-existing DW structures. Here we present further MFM data and micromagnetic simulations showing the controlled movement of DWs using lower magnetic tip charge values. Note that the overall magnetic charge of a 360\deg DW is zero as its composite 180\deg DWs carry equal and opposite charge. However the 180\deg DW closer to the tip-charge will experience a stronger interaction with the tip, resulting in a net force on the 360\deg DW and thereby facilitating tip-mediated motion.

Figure \ref{mfm-DW-drag} shows a sequence of MFM images of a 150 nm wide wire of radius of curvature $10~\upmu$m. The sequence depicts the same initial LM imaging a) and HM injection b) processes as described in the injection section above, the black arrow overlaid in b) indicates the final position of the HM tip prior to retraction. However, here the nanowire material is Py rather than Co. As Py is a softer magnetic material than Co, the LM tip previously used for imaging injected DWs instead achieves spatial manipulation of injected magnetic charges. Note that MFM images are performed in a raster fashion with each subsequent horizontal line measured at a later time. This allows the imaging of dynamic processes as seen below. Figure \ref{mfm-DW-drag} c) shows a scan moving from top to bottom of the image (direction indicated by the overlaid grey arrow) and shows the LM tip collecting a DW at $2~\upmu$m from the HM retraction point via an attractive magnetostatic interaction between tip and DW magnetic charges. The DW charge then follows the motion of the tip down the wire before stopping a few $\upmu$m from the bottom of the image, potentially due to pinning at a defect. Figure \ref{mfm-DW-drag} d) is a subsequent scan moving up the wire. We observe a magnetic charge being collected by the tip close to its final position in c) before moving with the tip up the length of the wire. To rule out the possibility that our contrast shows the wire becoming magnetised out of plane rather than a dynamic image of a moving DW we retracted the tip to $200~\upmu$m above the sample where the tip-sample interaction is negligible before moving the tip to the bottom of the wire. We then perform a second upwards scan, shown in e). As expected, the wire remains in an in-plane state. No magnetic charges are observed until $1~\upmu$m from the tip-retraction point at the top of d) where the charge is again collected by the tip. This shows we have successfully moved an injected magnetic charge through a nanowire and are able to deposit it at a desired position before re-collecting later if desired. This feature allows for the flexible re-configuration of a device's magnetic charge distribution with an accuracy of 1-2 $\upmu$m (the distance at which charges are attracted to the tip), a considerable advantage which is not achievable using existing DW injection techniques.

Here we have used a LM-tip to achieve spatial manipulation in Py. The same process is observed in Co wires using slightly stronger commercially available `normal' moment (NM) MFM tips.

\subsection*{Micromagnetic simulation of spatial manipulation}

\begin{figure}[tbp]
	\centering
	\includegraphics[width=15cm]{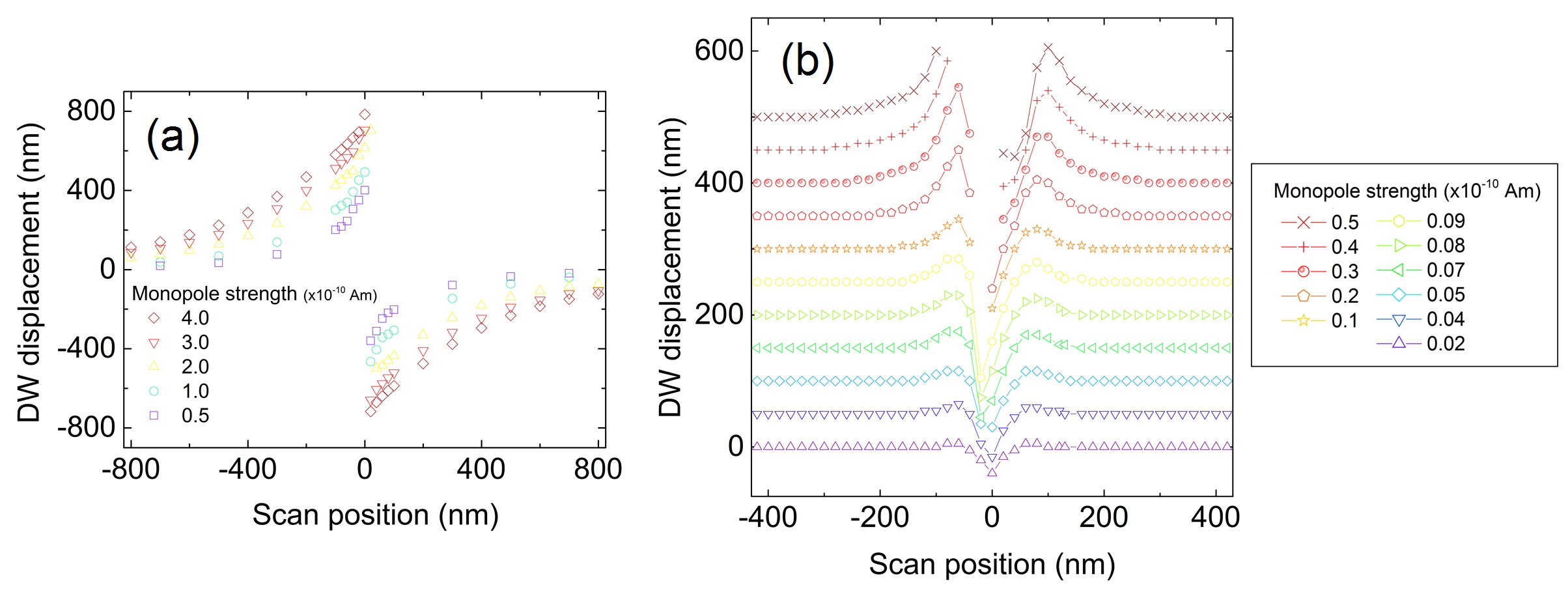}
	\caption{Spatial manipulation of a) an up-chirality transverse head-to-head domain wall and b) a clockwise-chirality 360\deg domain wall in a 100 nm wide nanowire by a passing monopole-like magnetic charge. Lines representing different monopole strengths in b) have been separated along the y-axis for ease of viewing, all monopole strengths give a DW displacement of zero at scan positions of $\pm$ 425 nm. Missing points in b) denote a destruction of the existing 360\deg DW by the magnetic charge. A scan position of 0 indicates that the magnetic charge crosses the wire directly at the location of the existing DW. The magnetic charge is suspended 5~nm above the wire.}
	\label{dw_manipulation}
\end{figure}

Figure \ref{dw_manipulation} plots the displacement of a DW along a wire following an interaction with a moving magnetic charge which crosses the wire at varying lateral distances from the DWs initial position. The displacement of the 180\deg DW is found by measuring the change in $M_x$ for the wire whilst the position of the 360\deg DW is found from the final micromagnetic configuration by taking the point of maximum $M_y$ magnetisation component along the nanowire axis. A video of the time evolution of the spatial manipulation process is included in the supplementary information.

The 180\deg DW has a magnetic charge which is repelled from the moving magnetic charge representing the tip (an oppositely magnetised tip or opposite polarity domain wall would lead to an attractive interaction). When the scan is near the DW location, the DW can be displaced by distances up to 800~nm. As the point at which the moving charge crosses the wire moves further from the DW, the magnitude of this displacement decreases. For the 360\deg DW structure, while a uniform magnetic field cannot induce a net force on the DW (due to opposite forces generated on each composite magnetic charge as described above) the localised field from the tip has a stronger influence on the closest of the two magnetic charges, leading to a weak displacement effect. This effect is evident in figure \ref{dw_manipulation} b) where tip-crossing operations performed to the left and right side of the DW both result in displacements in the positive x-direction. Again, the propagation distance decreases when the moving magnetic charge is further from the DW structure. This behaviour holds for scan positions greater than around the nanowire's width away from the existing DW, however below this limit different dynamics are observed. The central region about the y-axis in figure \ref{dw_manipulation} b) shows maximum positive displacements either side of zero, between which negative displacements are observed. The negative displacement reaches a maximum magnitude close to zero with a slight asymmetry caused by the moving magnetic charge experiencing attractive and repulsive interactions respectively to the opposite polarity constituent 180\deg DWs. The asymmetry is reversed about the y-axis by switching the polarity of the moving magnetic charge or the whole 360\deg DW. The shape of the central region of figure \ref{dw_manipulation} can be understood as the 360\deg DW experiencing a point of minimum energy when the constituent 180\deg DW of opposite polarity to the moving magnetic charge lies directly under the moving charge. At tip-crossing points within the finite width of the 360\deg DW, the DW will shift its position to reach this minimum energy point. The 180\deg DW with a matching polarity to the tip will move away while the opposing polarity 180\deg DW will approach the tip-charge, allowing for motion of the 360\deg DW in both directions along the wire. For moving magnetic charge strengths above $9\times{10}^{-12}$ Am the field of the moving charge can cause the $\pm 1/2$ topological edge defects on the near side of the wire to become unbound, leading to a collapse of the 360\deg DW as described above and a reversion to a uniformly magnetised state. These events occur when the moving charge crosses the wire close to the DW and are represented in figure \ref{dw_manipulation} by missing points in the central region. Experimentally, this DW deletion behaviour was not observed while using LM-MFM tips and as such we believe the spatial manipulation operations demonstrated are in the sub-$9\times 10^{-12}$ Am regime. 

\section*{Discussion}

In conclusion, we have performed experimental and micromagnetic studies on the interactions between a moving magnetic charge and a ferromagnetic nanowire with in-plane anisotropy. The spins in the nanowire experience the localised monopole-like field of the moving charge, leading to a change in spin orientation. With the movement of this charge over boundaries at the nanowire edges we show how topological defects can be introduced to a nanowire, creating micromagnetic structures such as 360\deg DWs. Our experimental work shows this process can be achieved using the magnetically charged tip of an MFM. 

The methodology described here serves to enhance the degree and sophistication of control available to researchers working on nanostructured magnetic systems, representing a significant addition to the existing toolbox of methods in its flexibility and ease of use. Particularly notable is that it does not build incrementally on existing electrical-stripline or global-magnetic-field based protocols as in the majority of recent advances, instead providing a substantially novel approach. The method allows for simple chirality control and spatial manipulation of injected 360\deg DWs, as well as the removal of the spatially fixed injection points and constraints on device architecture inherent in existing techniques. It is also possible to inject several 360\deg DWs at points along the same wire or perform injection across arrays of parallel wires using a single scan. The added spatial flexibility opens up a host of previously inaccessible device designs, creating promising avenues for future work.

\section*{Methods}

\subsection*{Structures and fabrication}

Nanostructures were fabricated using electron beam lithography followed by a liftoff process with permalloy (nominally Ni$_{81}$Fe$_{19}$) and Co deposited by thermal evaporation onto Si/SiO$_2$ substrates. 16~$\upmu$m long nanowire structures with widths of 80~nm to 150~nm and 10~nm thickness were prepared. Straight and curved nanowires with a 10~$\upmu$m radius of curvature were fabricated.

\subsection*{Magnetic force microscopy}

Measurements of local magnetisation states were performed using MFM with both HM and LM tips, magnetised out-of-plane relative to the nanowires. The contrast observed in MFM images provides a direct measure of magnetic charge \cite{Hubert-MFM-charge} and is thus ideal for our needs. AFM/MFM was performed in two distinct modes, injection and imaging/manipulation, using HM and LM tips respectively. The MFM system (Dimension 3100) was operated in an interleave lift mode where each raster line of the image consists of a preliminary atomic force microscopy (AFM) trace-retrace line followed by an MFM trace-retrace line with the tip raised to a specified lift-height (50-200 nm) above the sample. The injection and spatial manipulation processes occur during the AFM line, the raised-tip MFM line is used solely for imaging purposes. The AFM and MFM lines may be performed independently of one another to allow injection/spatial manipulation without simultaneous imaging or a pure imaging process where the magnetisation state of the nanostructure is left undisturbed. Scans were performed with the tip moving perpendicular to the wire length. 

With the exception of the AFM injection scan leading to the 360\deg DW state shown in figure \ref{mfm_inj-series} d), all scans (both AFM and MFM) were operated in a tapping mode where the tip-height is oscillated at the resonant frequency of the tip's cantilever. The injection scan corresponding to figure \ref{mfm_inj-series} d) was performed as a single AFM scan line (without an interleaved MFM line) using an HM tip in contact (non-tapping) mode. In contact mode the tip-height is not oscillated and instead tracks the topography of the sample surface at a near-constant tip-sample separation of {\raise.17ex\hbox{$\scriptstyle\mathtt{\sim}$}}2-5 nm. The HM and LM tips have moments and stray fields of {\raise.17ex\hbox{$\scriptstyle\mathtt{\sim}$}}$5\times{10}^{-13}$ emu, 690 Oe and {\raise.17ex\hbox{$\scriptstyle\mathtt{\sim}$}}$3\times{10}^{-14}$ emu, 320 Oe respectively\cite{jaafar2008calibration}, with stray fields measured at a typical AFM tip-sample separation ({\raise.17ex\hbox{$\scriptstyle\mathtt{\sim}$}}2-5 nm) away from the tip apex. 

\subsection*{Micromagnetic simulation}

Further insight into the magnetisation dynamics of the nanowires and their interactions with the localised magnetic field associated with the tip was obtained by performing a series of micromagnetic simulations using the object-oriented micromagnetic framework (OOMMF)\cite{Oommf}. Typical micromagnetic parameters for permalloy were used, i.e. saturation magnetisation, $M_S = 860 \times 10^{3}$~A/m, exchange stiffness, $A = 13 \times 10^{-12}$~J/m, zero magnetocrystalline anisotropy and a Gilbert damping parameter, $\alpha = 0.01$. The point probe approximation (that at small tip-sample separations an MFM tip may be described by a point monopole moment\cite{Phys.Lett.A_137_475, JAP_86_3410}) was used. This approximation is widely used in MFM simulations and previous work analysing systems with similar dynamics has shown that both dipolar and monopolar simulated tip fields induce the same tip-localised magnetisation vortices in thin magnetic films \cite{magiera2012magnetic,magiera2014magnetic}.

The simulated nanowires were 10~nm thick with widths ranging from 60~nm up to 150~nm  and were divided into $5 \times 5 \times 10$~nm cells. A semi-infinite wire was investigated using a 2~$\mu$m wide simulation window where the demagnetisation effects from the wire ends were corrected for by the inclusion of plates of fixed magnetic charge at the nanowire ends.\cite{IEEE.Trans.Magn_33_4167} 

The field from the MFM tip was modelled as a single magnetic charge, $q_{T}$, producing a radial field $H = \frac{1}{4\pi} \frac{q_{T}}{r^2}$ at a distance $r$ from the charge. During the simulation this magnetic charge moved perpendicular to the nanowire axis in 1~nm steps every 10~ps representing a velocity of 100~m/s.  This speed is faster than the velocities of $\sim 10^{-4}~m/s$ investigated experimentally. However, the simulated speed is still below those associated with exciting any precessional spin modes and is believed to be reasonable in this case. The position at which the magnetic charge crossed the nanowire was varied about the centre of the simulation window at a fixed height $h$ above the surface. Simulations were initialised with the magnetic charge 300~nm away from the nanowire in the plane of the wire to avoid a discrete jump in applied field at the wire on starting the simulation. Following the magnetic charge interaction with the wire, the wire was allowed to relax to an energetically stable state in zero field to obtain its final configuration.

The initial magnetisation of the system was prepared either uniformly magnetised along the wire axis or in an energetically minimised state obtained from a prior simulation in which a transverse 180\deg or 360\deg DW was located at the centre of the nanowire. 

\section*{Author contributions}

JCG, DMB and WRB conceived the experiment, DMB and JCG fabricated the samples and performed micromagnetic simulations, JCG performed the experimental measurements. JCG, DMB, WRB and LFC were responsible for drafting and revising the manuscript. LFC was responsible for key discussions and critical reading of the manuscript. 

\section*{Acknowledgements}

This work was supported by the Engineering and Physical Sciences Research Council [grant number EP/G004765/1] and the Leverhulme Trust [grant number RPG 2012-692] to WRB and supported by the Engineering and Physical Sciences Research Council [grant number EP/J014699/1] to LFC.

\section*{Data statement}

Data requests should be addressed to dataenquiryexss@imperial.ac.uk.

\section*{Competing interests}

The authors declare no competing financial interests.

\end{document}